\documentclass[10pt]{article}
\usepackage{upgreek}
\usepackage[mathscr]{euscript}
\usepackage{amsmath,environ}
\usepackage{amsfonts} 
\usepackage{amssymb}
\usepackage{mathrsfs}
\usepackage{graphicx,psfrag,color}
\usepackage{graphicx,amsmath,amssymb}
\usepackage{dcolumn}
\usepackage{bm}
\usepackage{graphicx,psfrag,color}
\usepackage{xcolor}
\usepackage[colorlinks=true,linkcolor=brown, citecolor=blue, urlcolor=purple]{hyperref}
\usepackage[top = 1.5in, left = 1in, bottom = 1in, right=1in]{geometry}

\usepackage[firstinits=true, parentracker=true, 
doi=true, isbn=false, url=true, natbib=true,
eprint=false,sorting=nyt,]{biblatex}

\begin{document}
\large

\title{Synchronization of globally coupled oscillators without symmetry in the distribution of natural frequencies\thanks{This work was completed in the period Oct 2004-Jan 2005 as part of the first author's postdoctoral stint at Cornell University. The author thanks Steven H Strogatz (strogatz@cornell.edu) for suggesting the problem. The analysis and results provided in section III of this paper were carried out, during this period, independent of the exposure to the results of \cite{Sakaguchi1986} (see also \cite{daido3,crawford9}). Several works have appeared related to the similar issues after this period, for instance \cite{
Basnarkov2008,
Petkoski2013,Terada2017}. Since the manipulations and observations presented in this paper are still anew it was decided by the authors to submit for publication.}}
\author{Basant Lal Sharma\thanks{(bs72@cornell.edu) Department of Theoretical and Applied Mechanics, Cornell University, Ithaca, NY 14853. 
Present address: {(bls@iitk.ac.in) Department of Mechanical Engineering, Indian Institute of Technology Kanpur, Kanpur, U. P. 208016, India}}}%

\date{Jan 6 2005}

\maketitle

\begin{abstract}
The collective behavior in a population of globally coupled oscillators with randomly distributed frequencies is studied when the natural frequency distribution does not possess an even symmetry with respect to the average natural frequency of oscillators. We study the special case of absence of symmetry induced by a group of scaling transformations of the continuous distribution of frequencies. When coupling between oscillators is increased beyond a critical threshold favoring spontaneous synchronization, we found that the variation in the velocity of the traveling wave depends on the extent of asymmetry in the natural frequency distribution. In particular for large coupling this velocity is the average natural frequency whereas at the onset of synchronization it corresponds to the frequency where the Hilbert Transform of the frequency distribution vanishes.
\end{abstract}

\section*{Introduction}
Populations of coupled oscillators arise in many branches of science and technology. For example phase-locked arrays of lasers \cite{wangwinful}, Josephson junctions \cite{hadleybeasley} and relativistic magnetrons \cite{benfordwoo}; biological rhythms in the heart \cite{winfree1, winfree2, glassmackey, grasmanjansen, michaels, jalife}, nervous system \cite{winfree1, winfree2, glassmackey, grasmanjansen, pavlidis, cohenholmes, kopell1, kopell2, traub}, intestine \cite{winfree2, kopell1} and pancreas \cite{sherman}; synchronous flashing of fireflies \cite{buck}, the chirping of crickets in unison \cite{walker}, and the mutual synchronization of menstrual cycles in groups of women \cite{mcclintock, russellswitz}; and recent applications of microelectronic mechanical systems \cite{hoppensteadt}.

The time dependent behavior of a large number of oscillators which are coupled together through a suitable mechanism lies at the intersection of statistical mechanics and nonlinear dynamics. During last four decades much work has been to solve many puzzles involving collective behavior of interacting oscillators. These results display the colorful nature of natural phenomena that arise due to interactions between self sustaining systems possessing stable limit cycles. Starting with mathematical modeling and emphasis on universal application of collective synchronization due to Winfree \cite{winfree1}, there were major advances in analysis made by Kuramoto \cite{kuramotob, kuramoto1} and later by Crawford \cite{crawford1, crawford2, crawford3, crawford4, crawford5, crawford6, crawford7, crawford8, crawford9}. A recent review by Strogatz \cite{strogatzr} narrates this interesting story.

Although the assumptions in mathematical models for these natural systems are often too restrictive, they lead to very illuminating conclusions. One such assumption has been the symmetry of the frequency distribution of the coupled oscillators. We drop this assumption of an even symmetry in this paper. Besides other reasons, the presence of an even symmetry allows rigorous work \cite{crawford2, crawford4, crawford9, mirollo, strogatz1} in the analysis. The critical values derived in the third section of this paper have been given in a slightly different but more general form in \cite{crawford9}.

By conventional wisdom the collective behavior of oscillators corresponds to propagation of traveling waves with velocity equal to the average natural frequency but we find that this conclusion is limited to systems with strong coupling when the natural frequencies are not distributed symmetrically about the average frequency. When coupling is close to the value that corresponds to the onset of synchronization the velocity of traveling wave may differ considerably if the distribution of natural frequencies is far from possessing the even symmetry. 

A study of the general case, such as \cite{crawford9}, of asymmetric distribution may involve more sophisticated analysis than that employed in this paper. We introduce asymmetry in a very special manner. Motivated by scaling transformations in mechanics, we deform the distribution of natural frequencies of the oscillators by a non uniform stretching of the natural frequencies. This scheme involves a group of transformations parameterised by one variable that also provides a `measure' of the extent of asymmetry.

In the first section we present the basic formulation that was introduced by Kuramoto \cite{kuramotob} and later adopted in the continuum framework \cite{strogatz1}. In the second section we derive the values of critical parameters at the transition when the synchronized state becomes stable for stronger coupling between the oscillators. In the third section we present a special group of transformations that leads to loss of the even property of the natural frequency distribution. In the last section we discuss some future possibilities.

\section{Kuramoto model}
Let $\{\theta _{i}:\mathbb{R\rightarrow }\mathbb{R}\}_{i=1}^{N}$ describe
the set of phase angles of $N$ oscillators with a prescribed initial phase such that
\begin{equation}
\frac{d}{dt}\theta _{i}(t)=\omega_{i}+\frac{1}{N}\sum_{j=1}^{N}\Gamma
_{ij}(\theta _{j}(t)-\theta _{i}(t)),i=1,...,N,
\end{equation}
with all $\Gamma _{ij}(.)$ periodic with period $2\pi $ and the number of oscillators having natural frequency $\omega_{i}$ is equal to $N$ times $g(\omega_{i})$. Clearly $\sum_{j=1}^{N}g(\omega_{i})=1.$ In the Kuramoto model $\Gamma _{ij}(x)=K\sin(x),$ $K\geq 0$ is the coupling strength and describes the global coupling between the oscillators \cite{kuramotob}.

Let $\rho :\mathbb{R\times \mathbb{R}}^{+}\mathbb{\mathbb{\times R}\rightarrow }\mathbb{R}$ describe the probability density $\rho (\theta,t,\omega )$, with respect to phase $\theta$ of the continuum \cite{strogatz1} of oscillators with natural frequency $\omega$ at time $t$ and $\rho (\theta ,t,\omega)=\rho (\theta +2\pi ,t,\omega )$ so that
\begin{equation}
\int_0^{2\pi}d\theta \rho (\theta,t,\omega )=1
\label{def_rho}
\end{equation}
for all $t$ and $\omega.$
Let $v:\mathbb{R\times \mathbb{R}}^{+}\mathbb{\mathbb{\times R\rightarrow }R}$ be the mapping describing the angular velocity of an oscillator with natural frequency $\omega $ at time $t$ and at location $\theta $
\begin{equation*}
v(\theta ,t,\omega )=\omega +\int_{-\pi }^{\pi }d\theta ^{\prime
}\int_{-\infty }^{\infty }d\omega ^{\prime }\Gamma (\theta ^{\prime }-\theta
)\rho (\theta ^{\prime },t,\omega ^{\prime })g(\omega ^{\prime }),
\end{equation*}%
with $\Gamma $ periodic with period $2\pi.$ $g(\omega )$ is the probability density of
the continuum of oscillators with natural frequency $\omega $. In the Kuramoto model
\begin{equation}
\frac{\partial }{\partial t}\rho (\theta ,t,\omega )=-\frac{\partial }{%
\partial \theta }(\rho (\theta ,t,\omega )[\omega +Kr(t)\sin(\psi
(t)-\theta )]),
\label{kmc}
\end{equation}%
with the {\em order parameter} $r(t)e^{i \psi(t)}$ defined as
\begin{equation}
r(t)e^{i\psi (t)}=\int_{-\pi }^{\pi }d\theta ^{\prime }\int_{-\infty
}^{\infty }d\omega ^{\prime }e^{i\theta ^{\prime }}\rho (\theta ^{\prime
},t,\omega ^{\prime })g(\omega ^{\prime }).
\label{cc1}
\end{equation}
The equation (\ref{cc1}) is also called the {\em consistency condition}. Henceforth we study the Kuramoto model within continuum framework.

We are interested in traveling waves in a population of globally coupled oscillators. In a continuum of oscillators we look for traveling waves with a velocity, yet unknown, $\Omega_0$ in accordance with equations (\ref{kmc}) and (\ref{cc1}). Let $\rho (\theta ,t,\omega )=\rho (\theta -\Omega_{0}t,\omega )$. For convenience we consider the moving frame $\psi (t)=\Omega_{0}t$ and $r$ is a constant in (\ref{cc1}). Thus $\rho (\theta,\omega )$ satisfies the following equation%
\begin{equation}
\frac{\partial }{\partial \theta }(\rho (\theta ,\omega )(\omega -\Omega_{0}+Kr\sin(-\theta ))=0
\label{sseqn}
\end{equation}%
with the new consistency condition
\begin{equation}
r=\int_{-\pi }^{\pi }d\theta ^{\prime }\int_{-\infty }^{\infty }d\omega
^{\prime }e^{i\theta ^{\prime }}\rho (\theta ^{\prime },\omega ^{\prime
})g(\omega ^{\prime }).
\label{sscc}
\end{equation}
The solution of (\ref{sseqn}) is given by $\rho(\theta ,\omega)(\omega -\Omega_{0}-Kr\sin\theta )=C(\omega).$ One solution that satisfies (\ref{sscc}) is $\rho(\theta ,\omega)=1/2\pi$ corresponding to the incoherent oscillators when each oscillators moves at its natural frequency, independent of other oscillators. However, we also have another solution
\begin{equation}
\rho (\theta ,\omega )=\left\{
\begin{array}{lll}
\delta (\omega -\Omega_{0}-Kr\sin\theta ),& \left| \omega -\Omega_{0}\right| \leq Kr, \\
C(\omega)/\left| \omega -\Omega_{0}-Kr\sin\theta \right|,& \left| \omega -\Omega_{0}\right| >Kr,
\end{array}
\right.
\label{sol}
\end{equation}
and by definition (\ref{def_rho}) of $\rho (\theta ,\omega )$ it can easily be found that $C(\omega)=\frac{1}{2\pi}((\omega -\Omega_{0})^{2}-K^{2}r^{2})^{1/2}$. For given value of $K$ and a distribution function $g$ the solution exists only if it satisfies the consistency condition (\ref{sscc}) which can be further simplified as
\begin{eqnarray}
Kr^{2} &=&\int_{-Kr}^{Kr}d\omega g(\omega +\Omega_{0})\sqrt{K^{2}r^{2}-\omega ^{2}}\label{reqn1}\\
0&=&\int_{-\infty }^{\infty }d\omega g(\omega +\Omega_{0})\omega\notag\\
&&-\int_{|\omega|>Kr}d\omega \frac{|\omega|}{\omega}g(\omega +\Omega_{0})\sqrt{\omega^{2}-K^{2}r^{2}}\label{Aeqn1}
\end{eqnarray}%
The proof is straightforward calculation. Assuming $r\neq 0$ and using new variables we get
\begin{eqnarray}
\frac{2\pi }{K} &=&\frac{\pi }{2}\int_{-\pi }^{\pi }d\theta g(Kr\sin\frac{\theta }{2}+\Omega_{0})\label{reqn2}\\
&&+\int_{-\pi }^{\pi }d\theta \cos\theta\lbrack \frac{\pi }{2}g(Kr\sin\frac{\theta }{2}+\Omega_{0})\notag\\
0 &=&\frac{1}{4}Kr \int_{-\pi }^{\pi }d\theta \sin\theta g(Kr\sin\frac{\theta }{2}+\Omega_{0})\label{Aeqn2}\\
&&+\int_{\left| \omega \right|>Kr}d\omega g(Kr\omega +\Omega_{0})(\omega- \frac{|\omega|}{\omega} \sqrt{\omega^2-K^2r^2}).\notag
\end{eqnarray}

\section{Critical Values at Phase Transition}
As a reminiscent of phase transition in matter, there is a sharp transition from `stable' \cite{mirollo} incoherent behavior to a stable coherent motion when coupling between the oscillators becomes larger than a certain threshold coupling. This critical value of $K$, called $K_c$, gives a lower bound of $K$ for the existence of solutions of type (\ref{sol}). By assuming $r=0$ in (\ref{reqn2}), we get $K_{c}=2/\pi g(\Omega_{0}).$ But it is impossible to find $\Omega_{0}$ by this method and therefore it is unknown at this point. A method as described below, based on asymptotic expansion at the critical point, determines the form of $\Omega_{0}$ for a given frequency distribution $g$.

Let $K\approx K_{c}$ so that $r\approx0$. To solve for $r$ we consider the real part (\ref{reqn1} or \ref{reqn2}) of the consistency condition (\ref{sscc}) to obtain
\begin{equation}
\frac{1}{K}=\frac{1}{2}g(\Omega_{0})\pi +\frac{1}{16}K^{2}r^{2}g^{\prime
\prime }(\Omega_{0})\pi +o(r^{2})
\label{reqn}
\end{equation}
as $r\rightarrow0.$ Here it is important to note that $g^{\prime \prime }(\Omega_{0})$ may not
necessarily have a constant sign. Let $\mu =\frac{K-K_{c}}{K_{c}}$ and we can simplify (\ref{reqn}) further and obtain%
\begin{equation}
r\approx \frac{4}{\sqrt{\pi }}\frac{1}{K_{c}^{3/2}}\sqrt{\frac{\mu }{%
-g^{\prime \prime }(\Omega_{0})}}.
\label{rasym}
\end{equation}

To solve for $\Omega_{0},$ we need to solve the equation (\ref{reqn2} or \ref{Aeqn2}) corresponding the imaginary part of the consistency condition (\ref{sscc}) for $r\approx0$. If we assume that $g$ is smooth at $\Omega_{0},$ then we get
\begin{eqnarray*}
0&=&\int_{\left| \omega \right| >0}d\omega \frac{g(\omega +\Omega_{0})}{\omega }+\frac{4}{3}Krg^{\prime }(\Omega_{0})\\
&&+\frac{1}{4}K^{2}r^{2}\int_{\left| \omega \right| >Kr}d\omega \frac{g(\omega +\Omega_{0})}{\omega ^{3}}+o(r^{2}).
\end{eqnarray*}
and therefore we require%
\begin{equation*}
\int\nolimits_{-\infty }^{\infty }d\omega \frac{g(\omega +\Omega_{0})}{%
\omega }=0.
\end{equation*}
 We henceforth denote the solution of above equation for $\Omega_0$ by $\omega_{c}$. If $g(\omega)$ is an even function of $\omega$ then $\omega_c=0$ is a solution. By a change of variable, we have%
\begin{equation}
\int_{-\infty }^{\infty }d\omega \frac{g(\omega )}{\omega -\omega_{c}}=0.
\label{A0eqn1}
\end{equation}
Let us use the definition of Hilbert Transform%
\begin{equation*}
\mathcal{H}(g)(\Omega)=\frac{1}{\pi }\int_{-\infty }^{\infty }\frac{%
g(\omega )}{\Omega -\omega },
\end{equation*}
then the solution for the critical velocity $\omega_{c}$ of the traveling wave at $K=K_{c}$ coincides with the root of the Hilbert Transform of $g$ or
\begin{equation}
\mathcal{H}(g)(\omega_{c})=0
\label{A0eqn2}
\end{equation}
and as stated earlier the critical value for $K$ is
\begin{equation}
K_c=\frac{2}{\pi g(\omega_{c})}
\label{Kceqn}
\end{equation}

For large coupling constant $K$ one obtains from equation (\ref{reqn1})
\begin{equation}
r=\lim_{Kr\rightarrow\infty}\int_{-Kr}^{Kr}d\omega g(\omega+\Omega_0)\sqrt{1-\frac{\omega^2}{K^2r^2}}=1
\end{equation}
and using equation (\ref{Aeqn1}) and
\begin{equation}
\lim_{Kr\rightarrow\infty}\int_{|\omega|>Kr}d\omega g(\omega+\Omega_0)\sqrt{K^2r^2-\omega^2}=0
\end{equation}
for sufficiently rapidly decaying function $g$ it can be found that
\begin{equation}
\int_{-\infty }^{\infty }d\omega g(\omega +\Omega_{0})\omega=0
\end{equation}
so that $\Omega_{0}$ is the average frequency $\omega_{avg}$.

\section{Special Asymmetry in the distribution of Natural Frequencies due to Scaling Transformation}

Let the frequency distribution be defined by%
\begin{equation}
g(\omega )=\frac{2ab}{a+b}\left\{ 
\begin{array}{cc}
\tilde{g}(b\omega ), & \omega \geq 0, \\ 
\tilde{g}(a\omega ), & \omega <0,%
\end{array}%
\right.
\label{transfeqn}
\end{equation}%
with $\tilde{g}$ a typical 
(smooth) frequency distribution symmetric about $\omega =0.$ It is sufficient to consider the case $b=1$ and $a>1$. For $a>1,$ $\Omega_{0}>0,$ and for $a\rightarrow 1$ it is expected that $\omega_{c}\rightarrow 0$.

We analyze the case when $a=1+\varepsilon $ with $\varepsilon \ll 1$. $\omega_{c}$ is determined by
\begin{equation*}
\int_{-\infty }^{-\omega_{c}}d\omega \frac{\tilde{g}(a(\omega +\omega_{c}))%
}{\omega }+\int_{-\omega_{c}}^{\infty }d\omega \frac{\tilde{g}(\omega
+\omega_{c})}{\omega }=0.
\end{equation*}
Let
\begin{equation}
\omega_{c}=\alpha _{1}\varepsilon +\alpha _{2}\varepsilon ^{2}+o(\varepsilon^2),
 \label{Acapproxm}
\end{equation}
as $\varepsilon\rightarrow0$ with $a=1+\varepsilon.$ Then 
\begin{eqnarray*}
&&0+\varepsilon \lbrack \tilde{g}(0)+\alpha _{1}\int_{-\infty }^{\infty
}d\omega \frac{\tilde{g}^{\prime }(\omega )}{\omega }] \\
&&+\varepsilon ^{2}[\alpha _{1}\int_{-\infty }^{0}d\omega \frac{\tilde{g}%
^{\prime }(\omega )}{\omega }+\frac{1}{2}\alpha _{1}\int_{-\infty
}^{0}d\omega \omega \tilde{g}^{\prime \prime }(\omega )\\
&&+\alpha _{1}\tilde{g}%
^{\prime }(0)+\frac{1}{2}\alpha _{1}^{2}\int_{-\infty }^{\infty }d\omega 
\frac{\tilde{g}^{\prime \prime }(\omega )}{\omega }+\alpha _{2}\int_{-\infty }^{\infty }d\omega \frac{\tilde{g}^{\prime
}(\omega )}{\omega }] \\
&=&0,
\end{eqnarray*}%
so
\begin{equation}
\alpha _{1}=\frac{\tilde{g}(0)}{-\int_{-\infty }^{\infty }d\omega \frac{\tilde{g}^{\prime }(\omega )}{\omega }}, 
\label{a1}
\end{equation}
\begin{equation}
\alpha _{2}=-\frac{\alpha _{1}}{2}+\frac{\alpha _{1}^{2}}{2\tilde{g}(0)}\int_{-\infty }^{0}d\omega \omega \tilde{g}^{\prime \prime }(\omega).
\label{a2}
\end{equation}
We can also find asymptotic expansion of $K_c=(1/a+1)/\pi \tilde{g}(\omega_c)$ in terms of $\varepsilon$ so that
\begin{equation}
K_c\approx\frac{1}{\pi\tilde{g}(0)}(2-\varepsilon+(1-{\alpha^2_1}\frac{\tilde{g}^{\prime\prime}(0)}{\tilde{g}(0)})\varepsilon^2)+o(\varepsilon^2)
\label{Kcapproxm}
\end{equation}
as $\varepsilon\rightarrow0$.

To study the effect of large $a$, let $a=\frac{1}{\varepsilon }$ with $\varepsilon \rightarrow 0^{+},$ then $\omega_{c}$ is determined by
\begin{equation*}
\varepsilon \int_{-\infty }^{0}d\omega \frac{\tilde{g}(\omega )}{\varepsilon
\omega -\omega_{c}}+\int_{0}^{\infty }d\omega \frac{\tilde{g}(\omega )}{%
\omega -\omega_{c}}=0,
\end{equation*}%
the first integral is a non-singular integral and it is bounded for nice $%
\tilde{g}$, therefore as $\varepsilon \rightarrow 0^{+},$ we get the desired
result%
\begin{equation}
\int_{0}^{\infty }d\omega \frac{\tilde{g}(\omega )}{\omega -\omega_{c}}=0,
\label{Aclargea}
\end{equation}
which is independent of $a$ (or $\varepsilon $).

For large coupling, $K\gg1$, we have $\Omega_0=\omega_{avg}$ with
\begin{equation}
\omega_{avg}=2(1-\frac{1}{a})\int_0^\infty d\omega \omega \tilde{g}(\omega)
\label{Aavg}
\end{equation}
so that
\begin{equation}
\omega_{avg}\approx2(\varepsilon-\varepsilon^2)\int_0^\infty d\omega \omega \tilde{g}(\omega)+o(\varepsilon^2).
\label{Aavgapprox}
\end{equation}

\section{Three Examples}
We consider three special cases of the symmetric distribution function $\tilde{g}.$

\begin{figure}[ht!]
\includegraphics[width=\linewidth]{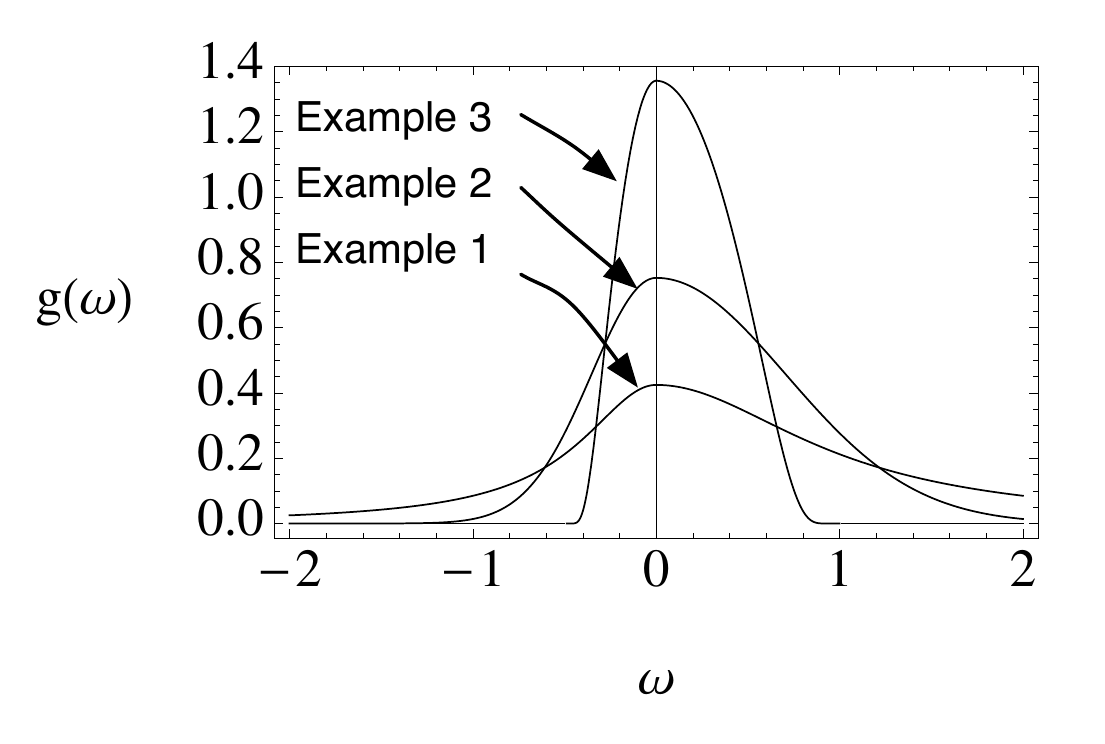}
\caption{Natural frequency distribution based on Example 1, Example 2, Example 3 and $b=1, a=2.$ Refer \eqref{Ex1}, \eqref{Ex2}, \eqref{Ex3}, and \eqref{transfeqn}.}
\label{figall_gwplt}
\end{figure}

\begin{figure}[ht!]
\includegraphics[width=\linewidth]{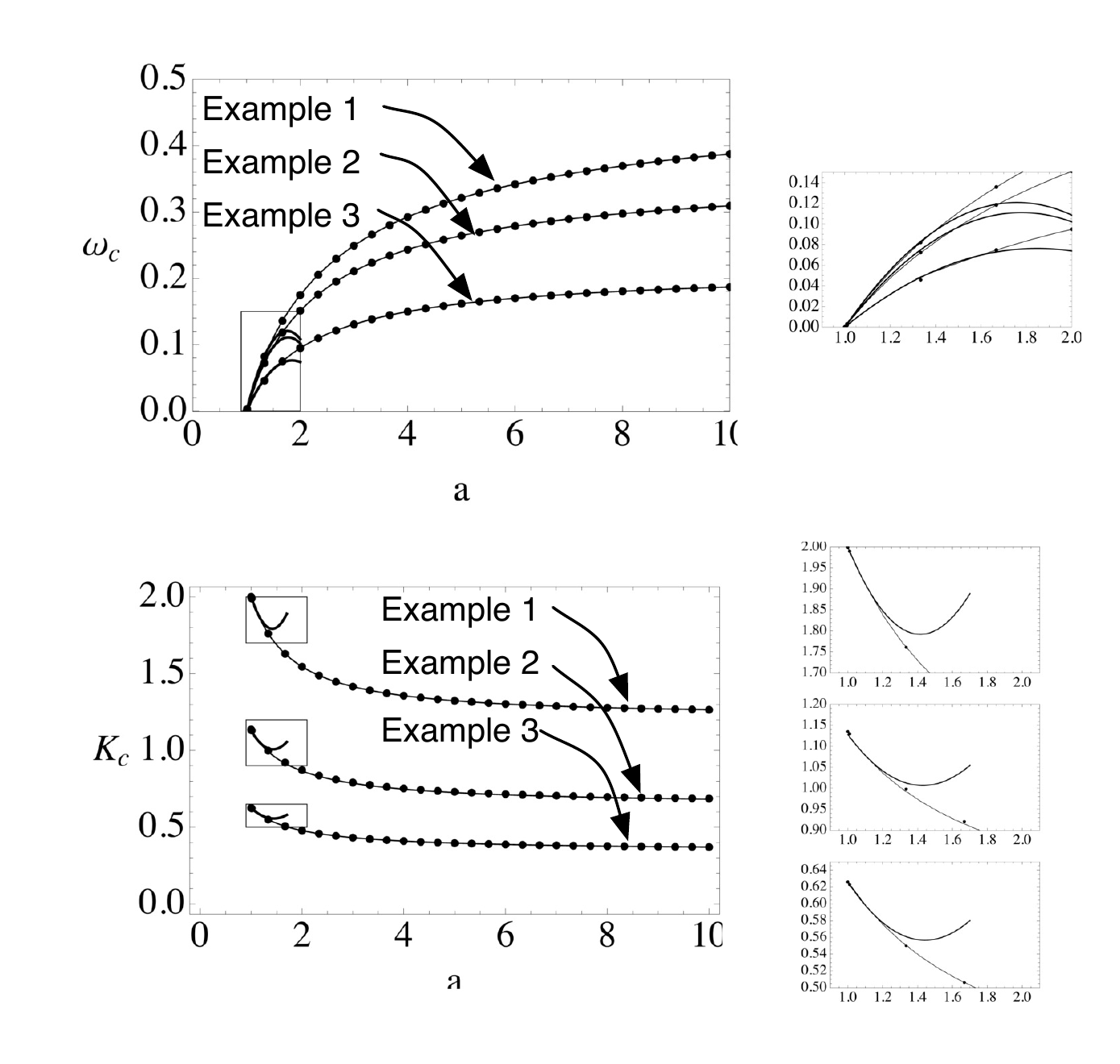}
\caption{Critical values for Example 1, Example 2, Example 3 as a function of $a>1$.}
\label{figall_WcKcaplts}
\end{figure}

\begin{description}
\item[Example 1]
The Lorentz distribution with a pathology that the average value of frequency is finite only in the symmetric case $a=1$
\begin{equation}
\tilde{g}(\omega )=\frac{1}{\pi }\frac{1}{1+\omega ^{2}}.
\label{Ex1}
\end{equation}
\item[Example 2]
The Gaussian distribution that permits oscillators with arbitrarily large frequency
\begin{equation}
\tilde{g}(\omega )=\frac{1}{\sqrt{\pi }}e^{-\omega^{2}} .
\label{Ex2}
\end{equation}
\item[Example 3]
A smooth distribution with compact support, so that for large enough (finite) coupling all oscillators are synchronized,
\begin{equation}
\tilde{g}(\omega )=\left\{
\begin{array}{cc}
Ce^{\frac{2}{\omega ^{2}-1}}, & \left|\omega\right|<1 \\
0, & \left| \omega \right| >1
\end{array}
\right.
\label{Ex3}
\end{equation}
with $C$ such that $\int_{-\infty}^{\infty}d\omega g(\omega)=1.$
\end{description}

\begin{figure*}[ht!]
\includegraphics[width=\linewidth]{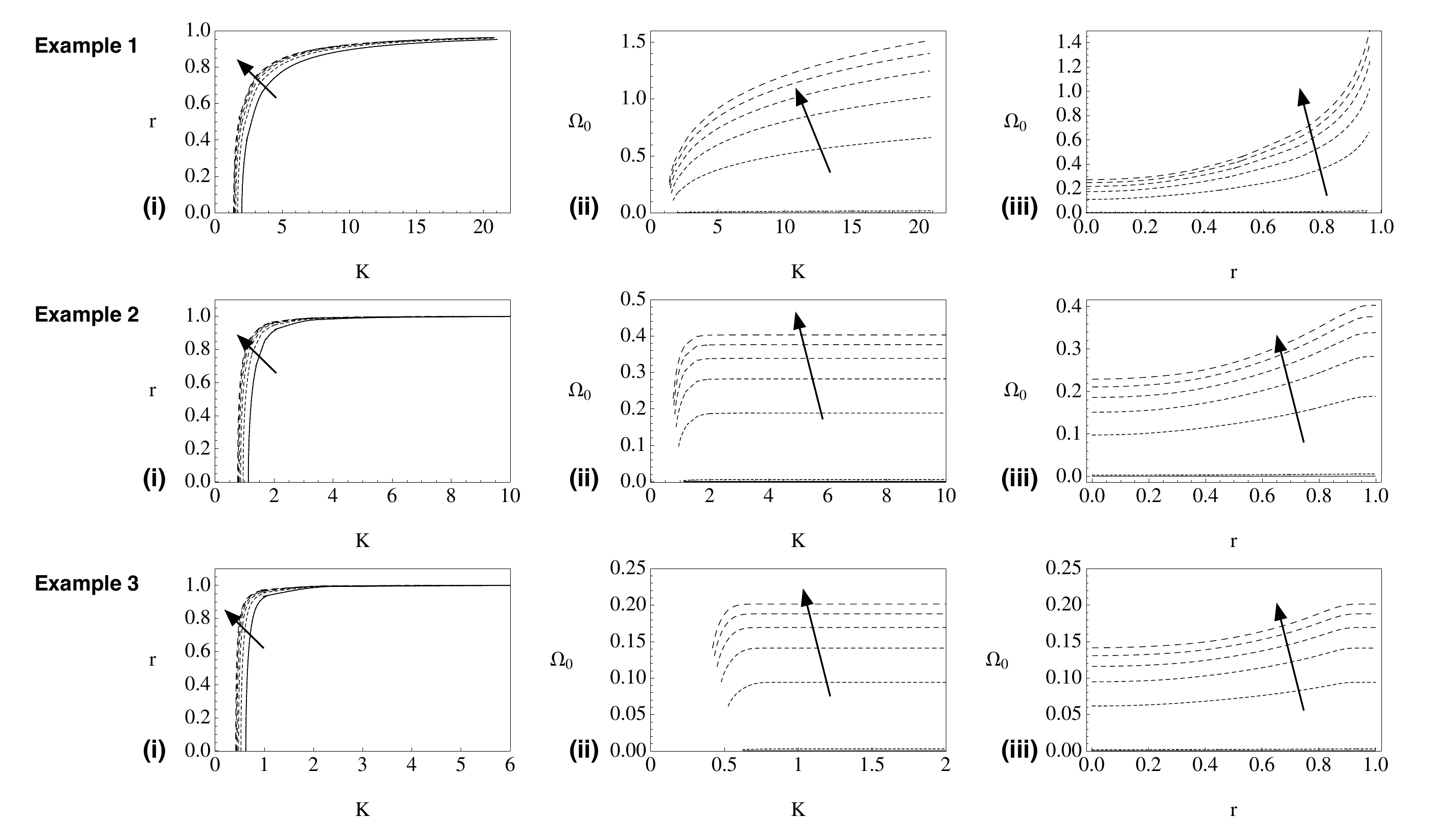}
\caption{An illustration of the relations between the order parameter $r$, coupling $K$, and velocity $\Omega_0$ for Example 1, Example 2, and Example 3, for different values of $a\in\{1,1.001, 1.01, 1.5, 2, 2.5, 3, 3.5\}$. Note that all plots have the same legend with dash length proportional to $a$ increasing in the direction of arrow.}
\label{Ex123}
\end{figure*}

For Example 1 the plot of the frequency distribution function $g$ after the scaling transformation (\ref{transfeqn}) of $\tilde{g}$ is shown in Fig. \ref{figall_gwplt}
for $a=2, b=1.$ By solving equations (\ref{reqn2},\ref{Aeqn2}) numerically and taking the limit $r\rightarrow0$ one can find $\omega_c$ and $K_c$. And by solving the equation (\ref{A0eqn2}) and using this $\omega_c$ in (\ref{Kceqn}) we can find these two quantities exactly. Further by using (\ref{Acapproxm}, \ref{a1}, \ref{a2}, \ref{Kcapproxm}), for Example 1 we get $\omega_{c}\approx\frac{1}{\pi }(\varepsilon -\varepsilon ^{2}+\frac{5}{6}\varepsilon ^{3})+o(\varepsilon ^{3})$ and $K_c\approx2-\varepsilon+(1+2/\pi^2)\varepsilon^2)+o(\varepsilon^2)$ as $\varepsilon\rightarrow0$. In Fig. \ref{figall_WcKcaplts} we compare these three ways of finding $\omega_c$ and $K_c$ and they agree very well.

When the coupling $K$ increases beyond the onset of synchronized state, we find that the velocity $\Omega_0$ approaches the average frequency $\omega_{avg}$. The specific trend is shown in Fig. \ref{Ex123}(ii) for different values of the `measure' $a$ of the asymmetry of the distribution $g$ as listed in the part (i) of the same figure. Note that $\omega_{avg}$ for Example 1 is infinite whenever $a\ne1.$ In part (i) of this figure we can observe that $r$ vs $K$ dependence doesn't change much with the values of $a$. Also (i) and (ii) have been combined into (iii) to see the relation between $r$ and $\Omega_0.$

\begin{figure}
\includegraphics[width=\linewidth]{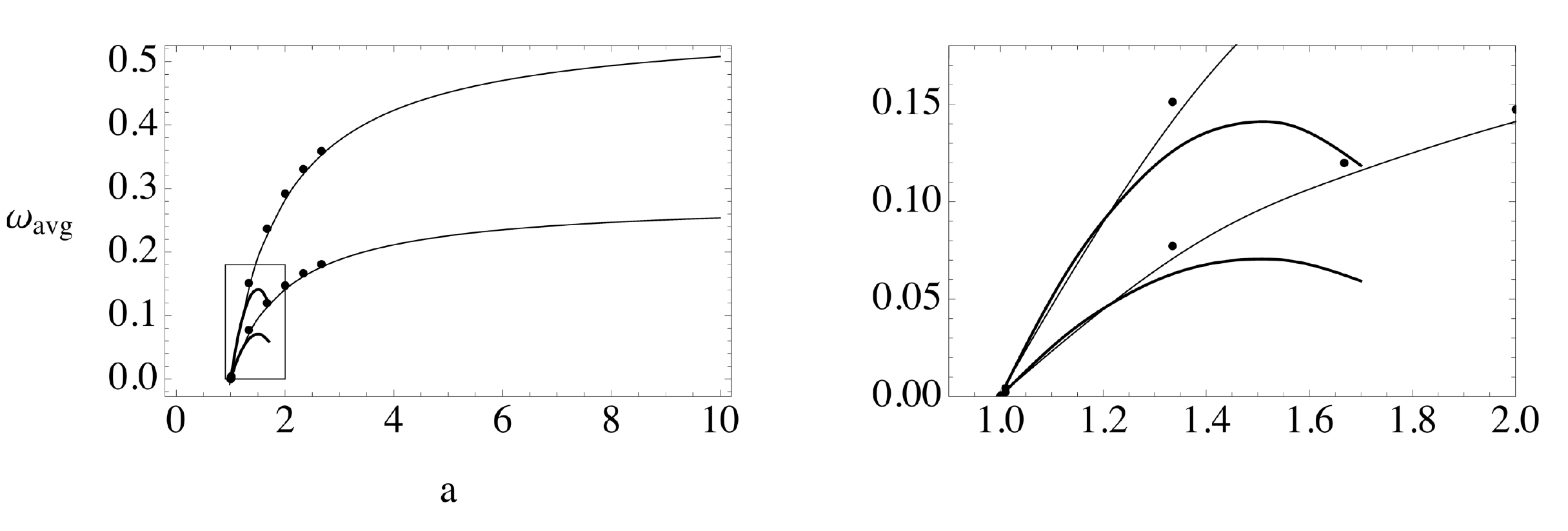}
\caption{Average frequency $\omega_{avg}$ for Example 2 and Example 3 as a function of $a>1.$}
\label{figall_Wavgaplts}
\end{figure}

For Example 2 the plot of $g$ is shown in Fig. \ref{figall_gwplt} for $a=2, b=1.$ We compare numerical results from (\ref{reqn2},\ref{Aeqn2}) and exact relations (\ref{A0eqn2}, \ref{Kceqn}) and asymptotic forms of (\ref{Acapproxm}, \ref{a1}, \ref{a2}, \ref{Kcapproxm}) in Fig. \ref{figall_WcKcaplts} and the agreement is again very good. In Fig. \ref{figall_Wavgaplts} is shown the dependence of $\omega_{avg}$ on the asymmetry parameter $a.$ As remarked for $\omega_c$ in the last part of the earlier section, as $a\rightarrow\infty$ we also obtain $\omega_{avg}\rightarrow constant$ different from the value given by (\ref{Aclargea}) for $\omega_c.$

For increase in coupling $K$ beyond $K_c$ the velocity $\Omega_0$ is shown in Fig. \ref{Ex123}(ii) for different values of $a$ in the part (i) of the same figure. Here $\omega_{avg}$ is finite and as $K$ becomes large enough $\Omega_0\approx\omega_{avg}$. Again, in part (i) of this figure we that $r$ vs $K$ dependence doesn't change much with the values of $a$. We also combine (i) and (ii) to observe the relation between $r$ and $\Omega_0$ in (iii).

Example 3 is very similar to Example 2 and the plot of $g$ is shown in Fig. \ref{figall_gwplt} for $a=2, b=1.$ We compare numerical results from (\ref{reqn2},\ref{Aeqn2}) and exact relations (\ref{A0eqn2}, \ref{Kceqn}) and asymptotic forms of (\ref{Acapproxm}, \ref{a1}, \ref{a2}, \ref{Kcapproxm}) in Fig. \ref{Ex123}. In Fig. \ref{figall_Wavgaplts} we can see the dependence of $\omega_{avg}$ on $a.$ As $K$ increases beyond $K_c$ the velocity $\Omega_0$ is shown in Fig. 
\ref{Ex123}(ii). As $K$ becomes large enough $\Omega_0\approx\omega_{avg}$. In Fig. \ref{Ex123}(i) we can again see that $r$ vs $K$ dependence doesn't change much with the values of $a$. We also combine (i) and (ii) to observe the relation between $r$ and $\Omega_0$ in (iii).

\section{Discussion and Open Problems}
The plots shown in Fig. \ref{Ex123} for Example 2 and Example 3 look very similar to each other. Also plots in Figs. \ref{figall_WcKcaplts}, 
and 
\ref{Ex123} as well as Fig. \ref{figall_Wavgaplts} 
suggest a ubiquitous scaling law perhaps associated with the specific transformation (\ref{transfeqn}). The parabolic profile in part (iii) of Fig. \ref{Ex123} 
for all three examples also warrants a asymptotic form for $\Omega_0$ near $r=0$ with a dependence on possibly local properties of $g$ at $\omega_c$ and specific to the asymmetry induced by the scaling transformation presented in this paper. The expressions similar to this for the general case may be more complicated and may dependent on more and finer properties of the natural frequency distribution $g.$

The asymptotic expansion of $\Omega_{0}$ near $\omega_{c}$ in powers of $\mu$ similar to the expression (\ref{rasym}) for $r$ may an interesting exercise. It is also an open question whether the results of analysis presented here can be extended to other models as well, i.e. with a non-sinusoidal coupling between the oscillators. The analysis of the stability of synchronized state with $K>K_c$ may involve slight modification of the work reported in \cite{strogatz1}. Also the connection between the approach presented here and described in \cite{crawford9} may be established in a future paper. 
\section{Acknowledgment}
We thank the National Science Foundation 
for the financial support during the period Oct 2004-Jan 2005. 

\renewcommand*{\bibfont}{\footnotesize}
\printbibliography
\end{document}